\documentclass[12pt,preprint]{aastex}
\usepackage{emulateapj5}
\usepackage{apjfonts}
\usepackage{epsfig}
\usepackage{natbib}

\newcommand{\etal}{et al.}

\newcommand{\feii}{\ion{Fe}{2}}

\def\gtrsim{\mathrel{\hbox{\rlap{\hbox{\lower4pt\hbox{$\sim$}}}\hbox{\raise2pt\hbox
{$>$}}}}}
\newcommand{\fopt}{\ensuremath{f_{4400~\mathrm{\AA}}}}

\newcommand{\fwhb}{\ensuremath{\mathrm{FWHM}_\mathrm{H{\beta}}}}
\newcommand{\halpha}{H\ensuremath{\alpha}}

\newcommand{\hbeta}{H\ensuremath{\beta}}

\newcommand{\kms}{km~s\ensuremath{^{-1}}}

\newcommand{\lum}{ergs s$^{-1}$}

\newcommand{\lledd}{\ensuremath{L_{\mathrm{bol}}/L{\mathrm{_{Edd}}}}}

\newcommand{\loiii}{\ensuremath{L_{\mathrm{[O {\tiny III}]}}}}

\newcommand{\mbh}{\ensuremath{M_\mathrm{BH}}}

\newcommand{\msigma}{\ensuremath{M_{\mathrm{BH}}-\sigmastar}}
\newcommand{\msun}{\ensuremath{M_{\odot}}}

\newcommand{\oiii}{[\ion{O}{3}]}
\newcommand{\psix}{\ensuremath{P_{\mathrm{6cm}}}}

\newcommand{\rkel}{{\emph R}}

\newcommand{\sigmastar}{\ensuremath{\sigma_{\ast}}}

\newcommand{\whz}{W~Hz\ensuremath{^{-1}}}

\def\lax{{$\mathrel{\hbox{\rlap{\hbox{\lower4pt\hbox{$\sim$}}}\hbox{$<$}}}$}}
\def\gax{{$\mathrel{\hbox{\rlap{\hbox{\lower4pt\hbox{$\sim$}}}\hbox{$>$}}}$}}

\slugcomment{To appear in {\it The Astrophysical Journal}.}
\shorttitle{Radio-quiet AGNs}
\shortauthors{GREENE, HO, \& ULVESTAD}

\begin{document}

\title{The Radio Quiescence of Active Galaxies with High Accretion Rates}

\author{Jenny E. Greene}
\affil{Harvard-Smithsonian Center for Astrophysics, 60 Garden St.,
Cambridge, MA 02138}

\author{Luis C. Ho}
\affil{The Observatories of the Carnegie Institution of Washington,
813 Santa Barbara St., Pasadena, CA 91101}

\and

\author{James S. Ulvestad}
\affil{National Radio Astronomy Observatory, P.O. Box 0, 1003 
Lopezville Road, Socorro, NM 87801}

\begin{abstract}

We present 6~cm Very Large Array observations of the Greene \& Ho
(2004) sample of 19 low-mass active galaxies with high accretion
rates.  This is one of the only studies of a uniform sample of
narrow-line Seyfert 1 (NLS1) galaxies with such high sensitivity and
resolution.  Although we detect only one source, the entire sample is
very radio-quiet down to strong limits.  GH10 was found to have a
radio power of 8.5 $\times 10^{21}$ \whz, and a ratio \rkel\ $\equiv
f_{6\mathrm{cm}}/f_{4400~\mathrm{\AA}}$ of 2.8.  The $3~\sigma$ upper
limits for the remaining nondetections correspond to radio powers from
$3 \times 10^{20}$ to $8 \times 10^{21}$ \whz and 0.47 $<$ \rkel\ $<$
9.9.  Stacking all nondetections yields an even stronger upper limit
of \rkel\ $\leq$ 0.27.  An assessment of existing observations in the 
literature confirms our finding that
NLS1s are consistently radio-quiet, with a radio-loud fraction of 0\%--6\%,
which is significantly lower than the 10\%--20\% observed in the
general quasar population.  By analogy with stellar-mass black holes,
we argue that AGNs undergo a state transition at \lledd\ $\approx
0.01$.  Below this value a radiatively inefficient accretion flow
effectively drives an outflow, which disappears when the flow turns
into an optically thick, geometrically thin disk, or a radiation
pressure-dominated slim disk at still higher \lledd.  

\end{abstract}


\section{Introduction}

Very little is known about the mass function of nuclear black holes
(BHs) with mass below $\sim 10^{6}$ \msun.  Until recently there were
only two secure candidates of intermediate-mass BHs in galactic
nuclei: NGC 4395 (Filippenko \& Ho 2003) and POX 52 (Barth \etal\
2004).  The tight correlation between host bulge velocity dispersion
and BH mass (the \msigma\ relation; Gebhardt \etal\ 2000; Ferrarese \&
Merritt 2000) suggests that BHs play an essential role in the
evolution of galaxies, and yet we know next to nothing empirical about
the starting conditions, or seeds, of supermassive BHs.  An
understanding of the low-mass end of the BH mass function may provide
one of the few observational constraints on seed BHs, at least prior
to the {\it Laser Interferometry Space Antenna} ({\it LISA}; Hughes
2002).  Furthermore, it remains unclear whether small galaxies,
without classical bulges, may host central BHs, and whether they obey
the same \msigma\ relation.  For these reasons, Greene \& Ho (2004)
performed a systematic search for such a population of
intermediate-mass BHs.  They used the First Data Release of the Sloan
Digital Sky Survey (SDSS;York \etal\ 2000; Abazajian \etal\ 2003) to
select a sample of broad-line active galactic nuclei (AGNs) with
virial mass estimates\footnote{Virial masses are calculated from the
relation between AGN luminosity and broad-line region radius
calibrated with reverberation-mapped AGNs (Kaspi \etal\ 2000) and a
measurement of velocity dispersion in the broad-line gas from the
full-width at half-maximum of the broad \halpha\ emission line (see,
e.g.,~ Greene \& Ho 2005).  If $\upsilon_{\mathrm{FWHM}}$ is the broad-line
gas velocity dispersion, and $L_{5100}$ is the AGN luminosity measured
at 5100 \AA, the luminosity-radius relation from Kaspi \etal\ (2000) yields 
\mbh\ = $4.82 \times 10^6~(L_{5100}/10^{44}~\mathrm{ergs~s^{-1}})^{0.7}
(\upsilon_{\mathrm{FWHM}}/1000~\mathrm{km~s^{-1}})$ \msun.} of \mbh\
$< 10^{6}$ \msun.  This sample of 19 galaxies represents the only
uniformly selected sample of intermediate-mass BHs in active galaxies.
Remarkably, these objects appear to obey the same \msigma\ relation as
that established for high-mass systems (Barth et al. 2005), suggesting
that a single physical mechanism operates over nearly 5 orders of
magnitude to maintain the observed relation.

This sample provides the opportunity to examine the broad spectral
energy distributions (SEDs) of BHs in a new mass regime.  As discussed
by Greene \& Ho (2004), the Eddington ratios [\lledd, where $L_{\mathrm{Edd}}
\equiv 1.26 \times 10^{38}$~(\mbh/\msun) \lum] of the sample are all close
to unity.  To begin to characterize the SEDs of this unique
sample, we obtained high-resolution 6~cm continuum observations using
the Very Large Array (VLA)\footnote{The VLA is operated by the National
Radio Astronomy Observatory, which is a facility of the National
Science Foundation, operated under cooperative agreement by Associated
Universities, Inc.}.  Our sample probes a poorly explored region of
parameter space, in terms of BH mass and Eddington ratio, and so may
provide new insights into the physical drivers of radio properties in
AGNs.

Throughout we assume the following cosmological parameters to calculate
distances: $H_0 = 100~h = 71$~\kms~Mpc$^{-1}$, $\Omega_{\rm m} = 0.27$,
and $\Omega_{\Lambda} = 0.75$ (Spergel \etal\ 2003).

\section{Observations and Data Analysis}

All observations were taken over 22.5 hours on 9 October 2004, when the
VLA was in A configuration (Thompson \etal\ 1980).  During the
observations only two antennas were excluded from the array, and the
weather conditions were good.  We observed at 4.860 GHz (6~cm; C
band) with a bandwidth of 
\psfig{file=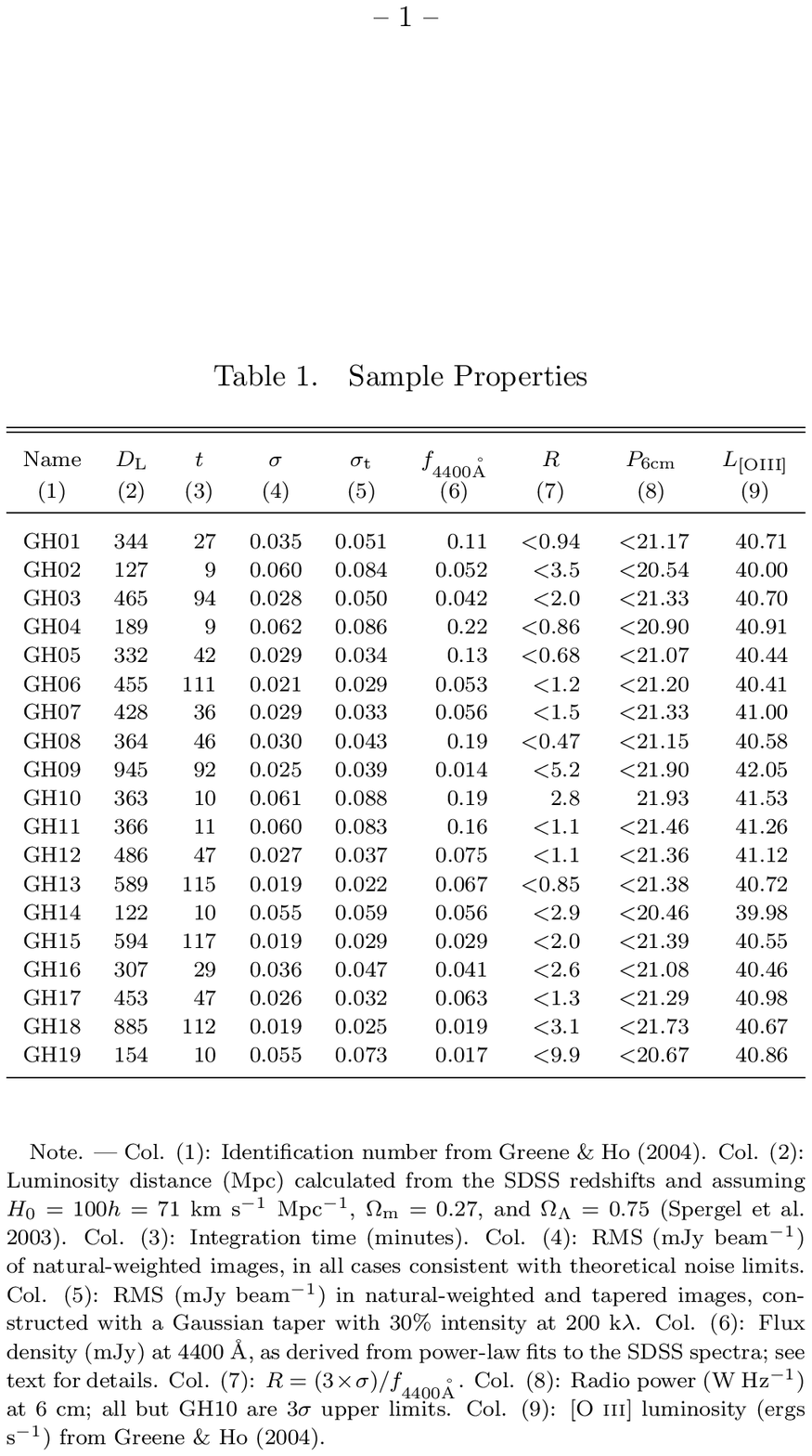,width=0.36\textwidth,keepaspectratio=true,angle=0}
\vskip -6mm
\noindent
50 MHz for each of two intermediate
frequencies separated by 50 MHz.  Integration times ranged from 10 minutes
to 2 hours, depending on the \oiii\ $\lambda 5007$ line luminosity of the 
sources given in Greene \& Ho (2004).
Using the relation between \loiii\ and radio power for radio-quiet
AGNs from Ho \& Peng (2001; shown in Fig. 1), we predicted a radio
power for each source, and thus the integration time required to reach
a $5~\sigma$ detection threshold (see Table 1), although we observed no
source for less than 10 minutes or longer than 2 hours.  For purposes of
phase calibration, observations of the target were alternated every
3--5 minutes with observations of a nearby (usually within $\sim
4$ degrees), bright radio source.  Our flux scale is tied to 3C 48, with
an assumed flux density of 5.4 Jy.  Flux-scale uncertainties are
dominated by uncertainties in the absolute flux of the phase
calibrators computed based on 3C 48 and are conservatively estimated
to be $\sim 5\%$.

Initial flux and phase calibration was performed within AIPS (Greisen
2003) on the observed visibilities, after removal of corrupted
records.  The visibilities were then Fourier transformed to create an
image of intensity on the sky.  Confusing sidelobes were removed using
the deconvolution algorithm CLEAN (H\"{o}gbom 1974) as modified by
Clark (1980).  Since this is a detection experiment, we use
``natural'' weighting to maximize our sensitivity.  The resulting maps
have a pixel scale of $0\farcs06$ and a typical synthesized beam of
$\Delta \theta \approx0\farcs6$.  The half power beam width (primary beam) of
the individual antennas is $\sim 9^{\prime}$, so we constructed images of this 
size to search for confusing sources in the field, which we found in only two
cases (GH07 and GH14).  The theoretical noise ranges from 0.02 to 0.06 mJy 
beam$^{-1}$. Since CLEAN can change the noise statistics in a 
source-free image, aside from the three objects with flux in the maps
(GH07, GH10, and GH14), we performed subsequent analysis directly on the 
``dirty'' maps.  In all cases we were able

\psfig{file=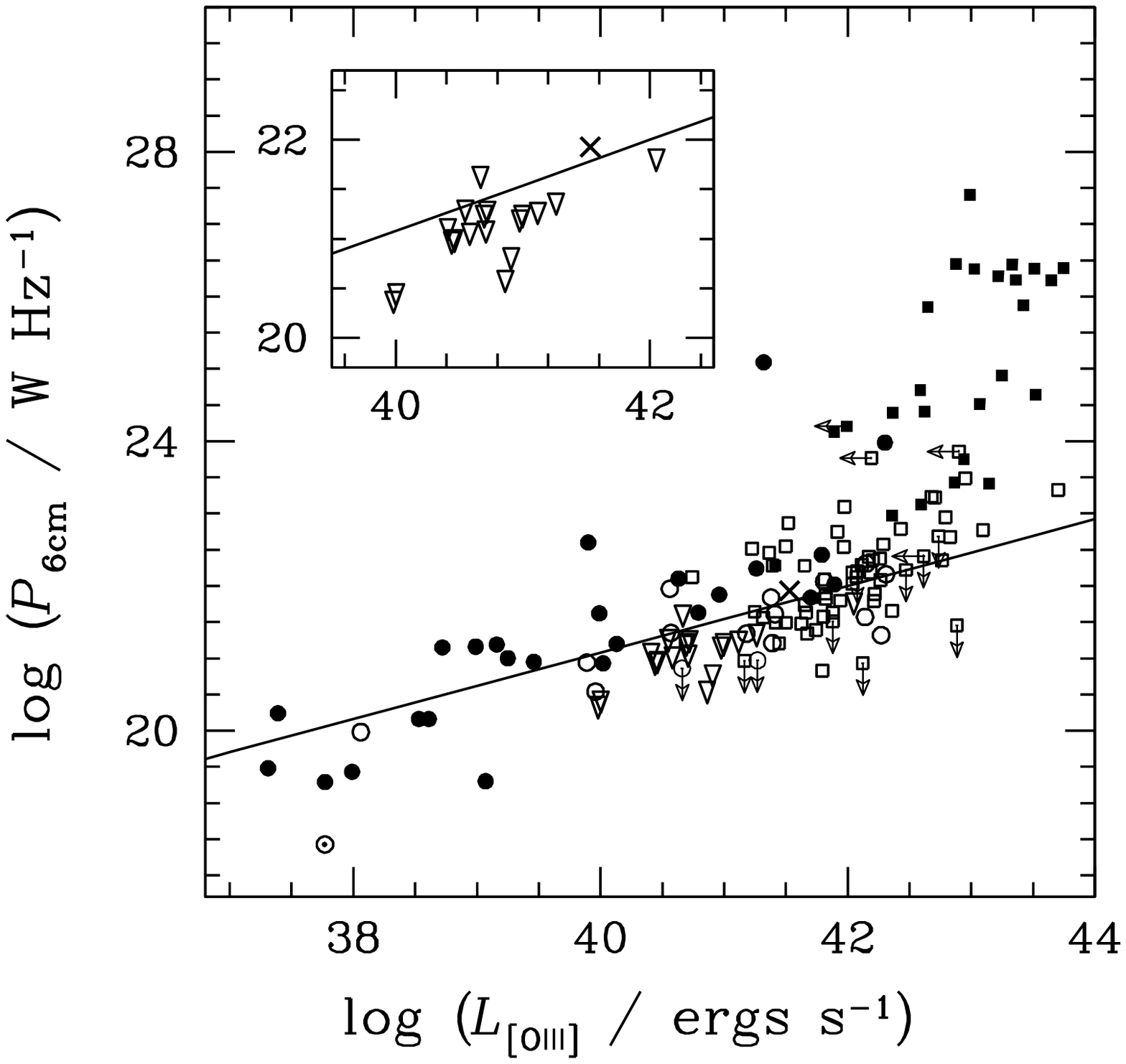,width=0.5\textwidth,keepaspectratio=true,angle=0}
\vskip -8mm 
\figcaption[]{ 
The trend of increasing radio power with
increasing [O {\tiny III}]~$\lambda 5007$ luminosity.  Upper limits
from this paper are shown as open triangles, while the GH10 detection
is shown as a cross.  For clarity, our data are plotted alone in the
inset.  The solid line represents the fit from Ho \& Peng (2001) to
radio-quiet Seyferts and PG quasars: \psix\ = (0.46 $\pm$
0.15)~$L_{\mathrm{[O {\tiny III}]}}$ + (2.68 $\pm$ 6.21).  The Ho \&
Peng sample is overplotted with filled (radio-loud sources) and open
(radio-quiet sources) symbols.  PG quasars are shown as squares, and
Seyferts are shown as circles.  NGC 4395 (adjusted to a distance of 4.2
Mpc; Thim et al. 2004) is highlighted as a semifilled circle.  See
text for details.  }
\vskip +3mm

\noindent
to achieve the
theoretical noise limit (see Table 1).  In the three cases where CLEAN
was required, we iterated until the ``clean components'' reached a
limiting flux of $3 \times$~RMS, where RMS is the theoretical noise
limit, and in the case of GH07 and GH14, clean boxes were placed around the 
distant confusing sources.  Finally, to investigate the possible
presence of more extended emission, we also constructed images with a
Gaussian tapering function that fell to 30\% power at 200 k$\lambda$,
which resulted in a typical synthesized beam of $\sim 1$\arcsec.  We
note that our minimum spacing is $\sim 11$~k$\lambda$, corresponding
to a maximum angular scale of $\leq 10$\arcsec.

\vspace{+0.3cm}
\section{Results}

Only one object in this study, GH10, was detected.  For the remaining
undetected targets, we computed $3~\sigma$ upper limits on their 6~cm
radio powers, which range from $3 \times 10^{20}$ to $8 \times
10^{21}$ W~Hz$^{-1}$, and are shown as a function of \loiii\ in Figure
1 (triangles).  The solid line is the \loiii-\psix\ relation from Ho
\& Peng (2001) that we used to calculate observing times for our
sample.  Clearly our measurements do not follow this relation (note
that the majority of the upper limits lie below the line, since our
integration times were designed to achieve $5~\sigma$ detections;
exceptions are objects that required more than two hours of
integration time).  In order to illustrate the probable cause of the
discrepancy, we include in Figure 1 the data from Ho \& Peng used to
derive our adopted \loiii-\psix\ relation.  Three samples were
included, to span the maximum range in optical nuclear luminosity ($-8
\leq M_{B} \leq -28$ mag).  The highest-luminosity sources are the 87
Palomar-Green (PG) quasars with $z < 0.5$ (Schmidt \& Green 1983),
while low-luminosity Seyfert galaxies are drawn from the Palomar
spectroscopic survey of nearby galaxies (Ho et al. 1995, 1997),
supplemented at intermediate luminosity by the sample of Seyferts
selected from the CfA redshift rurvey (Huchra \& Burg 1992).  The
radio observations for the CfA Seyferts are presented in Kukula \etal\
(1995), those for the Palomar Seyferts are from Ho \& Ulvestad (2001)
and Ulvestad \& Ho (2001a), and the PG quasars were observed by
Kellermann \etal\ (1989).  As shown in Figure 1, our upper limits are
lower than the typical PG quasars, whose radio powers range from $7
\times 10^{20}$ to $3 \times 10^{27}$ W~Hz$^{-1}$. [For reference the
3.6~cm radio power of the quasars in the Large Bright Quasar Survey
(LBQS) range from $10^{23}$ to $10^{28}$~\whz\ (Visnovsky et
al. 1992).]  They are more consistent with the distribution of radio
powers of the Seyfert nuclei ($\sim 10^{18}$ to $10^{25}$
W~Hz$^{-1}$).  Radio powers alone, however, are somewhat misleading,
since BHs of different mass have different (limiting) Eddington
luminosities. Complementary information is provided by the ratio $R
\equiv f_{6\mathrm{cm}}/f_{4400~\mathrm{\AA}}$, where ``radio-loud''
objects are conventionally defined as those with \rkel\ $\geq 10$
(Kellermann \etal\ 1989).  Radio-loud and radio-quiet objects are
represented as filled and open symbols in Figure 1, respectively.
Among quasars, the radio-loud fraction is $\sim 20\%$ for the PG
sample and $\sim 10\%$ for the LBQS sample.  In contrast, $\geq 60\%$
of the Ho \& Peng Seyfert sample are radio-loud, once the nuclear
emission is properly isolated.  

Now we investigate possible reasons
that the Ho \& Peng \loiii-\psix\ relation might overpredict the radio
power of our sources.  The relation was fit to the radio-quiet points
only, and since our entire sample is radio-quiet (GH10 has \rkel\ =
2.8, while the mean \rkel\ for the rest is $<$ 2.3), the discrepancy
cannot be explained by our choice of the optical-radio relation.
However, very few of the radio-quiet objects from Ho \& Peng have
\oiii\ luminosities less than $10^{41}$ ergs~s$^{-1}$, while the
majority of our objects do.  Either there is a break in the
\loiii-\psix\ relation at low \loiii, or the true slope of the
\loiii-\psix\ relation is steeper than that found by Ho \& Peng, due
to their limited dynamic range.  Actual detections of a sample such as
that presented here are required to distinguish between these two
possibilities.

\subsection{GH10}

GH10 is the only object in the sample detected by the Faint Images of
the Radio Sky at Twenty-cm Survey (FIRST; Becker et al.  1995) and is
the only object detected here.  Because of the relatively large beam
($\sim$ 5\arcsec) of FIRST and the low radio power of GH10 ($2.0
\times 10^{22}$ W~Hz$^{-1}$), Greene \& Ho (2004) argue that the 20~cm
emission from GH10 may originate from the host galaxy rather than the
AGN.  Given the much superior angular resolution of our current image
($\Delta \theta = 0\farcs57 \times 0\farcs42$), most of the radio
emission probably originates from the AGN itself (see Fig. 2).  Using
the task JMFIT within AIPS, we find that the image is marginally
resolved, with a deconvolved size of $0\farcs3 \times 0\farcs2$
($530~{\rm pc} \times 350~{\rm pc}$), a position angle of $9 \degr$,
and a total flux density of $0.7 \pm 0.1$ mJy.  Since the source is
slightly resolved, there is most likely some host galaxy contribution,
and so the peak flux density may give a better measure of the
unresolved AGN emission.  The peak flux density of $0.54 \pm 0.05$ mJy
corresponds to a radio power of $8.5 \times 10^{21}$~W~Hz$^{-1}$.
Using the 4400 \AA\ flux density from Table 1, we find an $R$ value of
2.8 for this source.  Since the source was also detected with FIRST,
we can estimate a crude spectral index $\alpha$, where $f_{\nu}
\propto \nu^{-\alpha}$.  This 
calculation is problematic because the
observations were nonsimultaneous and made with very different
resolutions.  In order to minimize the uncertainties resulting from
differing 
\psfig{file=GH10_BETTER.PS,width=0.4\textwidth,keepaspectratio=true,angle=-90}
\vskip +1mm
\figcaption[]{
The A-array image of GH10 at 6~cm.  The contours are $-3$, $-2$, 2, 3, 6, 10 
$\times$ 0.052 mJy beam$^{-1}$, which is the nominal RMS for the image.  The
beam size, shown in the lower-left corner, is $0\farcs57 \times 0\farcs42$.
The coordinates are offset from the nominal optical SDSS position of GH10.
}
\vskip 5mm

\noindent
resolutions, we use the FIRST peak flux density, 1.2 mJy beam$^{-1}$,
to find $\alpha$ = 0.62.  We emphasize that this is a very uncertain
value.

Unless otherwise stated, we omit GH10 from the following discussion.

\subsection{Upper Limits}

Given that all but one of the objects in our sample were
nondetections, we have devoted some care to deriving the most
stringent possible upper limits on the radio emission from these
sources.  We derive RMS values from our final images within a box
enclosing the inner quarter of the image, and our nominal upper limits
are simply 3 times this measured RMS.  At worst, the instrumental and
atmospheric phase for each antenna on a given calibrator vary by no
more than $\sim 20 \degr$ over a 5-minute period.  There are no
substantial phase changes even between calibrators in very different
regions of the sky; over the course of 18 hours of our 24-hour
observation, the phase calibration remained stable within $10 \degr$
for each antenna.  This implies a coherence loss of less than $1\%$
for our target objects, independent of the angular separation between
calibrator and target, leading to the conclusion that our cited upper
limits are accurate and quite robust.

While we were unable to detect any of our objects individually, we can
construct a single, deep image by coadding all the observations.  This
approach has been applied successfully with nondetections in FIRST of
SDSS (Glikman \etal\ 2004) and 2QZ (Wals \etal\ 2005) AGNs.  Stacking
is performed in the image plane.  Since the FIRST observations have
uniform sensitivity, they can be summed directly.  In our case,
because exposure times varied between observations, we give additional
weight to those with the highest sensitivity.  We compute a
weighted-average image using the task COMB within AIPS, where the
weights are simply $1/\mathrm{RMS}^2$ and the images are aligned to their
central pixel.  We expect all the sources to lie within the central
$0\farcs1$, since our positions depend on SDSS astrometry, which has
an accurary of $\leq 0\farcs1$ (Stoughton \etal\ 2001), and the
astrometric accuracy of the VLA phase calibrators is also $\leq 0\farcs1$
(Wilkinson \etal\ 1998; VLA calibrator manual\footnote{{\tt
http://www.vla.nrao.edu/astro/}}).  The individual maps used in the
stacking analysis were thus constructed with $0\farcs1$ pixels.  The
effective RMS of our stacked image is 0.0069 mJy beam$^{-1}$, which is
approximately one-third of our best RMS for a single exposure.  This
is roughly as expected, since we have increased the on-source exposure
time from a maximum of 2 hours to 16 hours.  We take our upper limit
as $3 \times$RMS and an average $f_{4400~\mathrm{\AA}}$ = 0.077 mJy
(from the values published in Greene \& Ho 2004; see Table 1) to find
an effective limit on the $R$ parameter of 0.27.  We are therefore
able to place a far more stringent upper limit for the sample ensemble
than for any individual source.

For completeness, we also derived upper limits on extended emission
from the host galaxies themselves, using lower-resolution, tapered
images with a synthesized beam $\Delta\theta$ = 1\arcsec.  Our upper
limits correspond to an average 6~cm radio power of $\sim 1.8 \times
10^{21}$~W~Hz$^{-1}$, which is roughly similar to the 1.4 GHz power of
normal (inactive) $L^*$ galaxies (Condon 1992).  Given that the
galaxies in this sample are relatively faint, $\sim 1$~mag fainter
than $L^*$ (Greene \& Ho 2004), this is not a very strong limit, but
it does indicate that the hosts are not experiencing any vigorous
starburst activity.  Assuming a radio spectral index of $\alpha =
0.7$, the inferred average 20~cm radio power of $4.2 \times
10^{21}$~W~Hz$^{-1}$ translates into an approximate star formation
rate of $\sim 2.5$ $M_\odot ~{\rm yr}^{-1}$ (Yun et al. 2001).

\subsection{Comparison with Narrow-Line Seyfert 1 Galaxies}

A natural comparison may be drawn between our sample and NGC 4395, the
prototypical AGN with an intermediate-mass BH.  NGC 4395 has been
imaged in the radio with the VLA (Moran et al. 1999; Ho \& Ulvestad
2001) and the Very Long Baseline Array (VLBA; Wrobel et al. 2001).  At
an assumed distance of 4.2 Mpc (Thim et al. 2004), the VLA
observations of Ho \& Ulvestad (2001) give a 6~cm power of $1.7 \times
10^{21}$~\whz, a 20~cm power of $3.5 \times 10^{21}$~\whz, and a
spectral index of $\alpha = 0.6$.  The 20~cm VLBA image has a diameter
less than 11~mas and a total power of $1.1\times 10^{21}$~\whz,
implying that some of the flux has been resolved by the VLBA.  Deeper
VLBA imaging of NGC 4395 is underway to search for a potential jet
component (J.~M.  Wrobel, J.~S.~Ulvestad, \& L.~C. Ho in preparation).
Using the \fopt\ measurement from Filippenko \& Ho (2003), we find
that NGC 4395 has \rkel\ $\approx$ 2.0.

More generally, the objects presented here technically belong to the
subclass of AGNs known as narrow-line Seyfert 1 galaxies (NLS1s;
Osterbrock \& Pogge 1985).  Formally, these are broad-line AGNs with
FWHM $\leq 2000$~\kms\ for the \hbeta\ line.  They also tend to have
low \oiii/\hbeta\ ratios, high \feii/\hbeta\ ratios, and prominent
soft X-ray excesses (e.g.,~Boller et al.  1996; but see Williams et
al. 2004).  This set of properties suggests that NLS1s are low-mass
BHs radiating near their Eddington limits (Pounds et al. 1995).  All
of our objects (as does NGC 4395) qualify as NLS1s based
on the \hbeta\ linewidth criterion, although their \feii\ and \oiii\
strengths cover a wider range than typical NLS1s (Greene \& Ho 2004),
and therefore comparisons with NLS1 properties are natural.

Anecdotally, NLS1s are thought to be radio-quiet as a class, although few 
statistical samples have been considered in the literature.  Ulvestad et al.
(1995) assembled new and published radio data for a total of 15 NLS1s, of 
which nine were detections.  We have gathered optical continuum luminosities 
for 11 of the Ulvestad et al. objects (excluding Mrk 291,
Mrk 957, IRAS 1509$-$211, and 1747.3+6836), using \hbeta\ flux and
equivalent width measurements from Osterbrock \& Pogge (1985)
when available, or else manually estimating the continuum fluxes from
spectra published in V{\'e}ron-Cetty et al. (2001).  Mrk 783, with \rkel\ = 
400, is the only radio-loud object in the sample.  As pointed out by Ulvestad 
\etal, their sample of NLS1s is by no means complete, making this statistic 
difficult to interpret.  

A number of other studies have used large radio surveys to estimate
the radio-loud fraction among NLS1s.  Zhou \& Wang (2002) collected
SEDs from the literature for a sample of 205 NLS1s in the
V{\'e}ron-Cetty \& V{\' e}ron (2001) catalog.  They defined a
subsample of 182 sources that fall within the footprint of the NRAO
VLA Sky Survey (NVSS; Condon \etal\ 1998).  Of these, 63 are detected
and 11 (6$\%$) qualify as radio-loud (\rkel\ $> 10$), although none
are ``very'' radio-loud (\rkel\ $> 300$).  Zhou \etal\ (2003) discuss
the radio properties of a sample of 175 NLS1s from the SDSS Early Data
Release (Stoughton \etal\ 2002).  It is unclear exactly what the
selection criteria were for this sample, but $\sim 6\%$ are detected
by FIRST, and two (1\%) of those can be deemed radio-loud.  Lastly,
Stepanian \etal\ (2003) presented SEDs from the literature for the
NLS1s in the Second Byurkanan Survey (Markarian et al. 1983), selected
to have \fwhb\ $< 2000$ \kms\ and \oiii/\hbeta\ $<$ 3.  Ten of their
26 objects have radio detections, and all are radio-quiet.  In all the
above cases, the fraction of radio-loud NLS1s seems to be low
(0\%--6\%) in comparison to the parent samples of normal, luminous
broad-line AGNs (10\%--20\%).  Unfortunately our ability to interpret the
nondetections is hampered by the use of relatively shallow surveys at
low resolution (e.g., NVSS or FIRST).  Nondetections may be radio-loud
or radio-quiet, while detections may contain significant contamination
from the host galaxy.  For instance, a typical NLS1 at $z=0.1$ with
\mbh\ = $10^6$~ \msun\ and \lledd\ = 1 will not be detected by FIRST
below a limiting \rkel\ of 17 [assuming a 1~mJy detection limit and a
flat ($\alpha=0$) radio spectrum].  Moreover, the definition of NLS1s
varies from study to study, with unknown consequences for the
underlying properties of each group.

A possibly more pertinent comparison with our sample comes from the PG survey, 
since it has available sensitive, high-resolution radio observations 
(Kellermann et al. 1989) and homogeneous optical spectrophotometric data 
(Neugebauer et al. 1987).  Among the 87 $z < 0.5$ sources, 18 meet the 
\hbeta\ linewidth criterion of NLS1s (Boroson \& Green 1992).  Their 6~cm 
radio powers range from 1.7 $\times 10^{21}$ to 2.5 $\times 10^{24}$ 
W~Hz$^{-1}$, and they are predominantly radio-quiet (94\%), with only a single 
radio-loud object (PG 1211+143).  The radio-quiet objects have \rkel\ from 0.2 
to 2.5, and a mean of 0.54.  Interestingly, the stacked upper limit of our 
sample is lower than the mean \rkel\ for the PG NLS1s.  

While the evidence summarized above suggests that NLS1s, as a class,
are exceptionally radio-quiet, it is worth noting that a few bona
fide radio-loud NLS1s have been identified in the literature (Siebert
\etal\ 1999; Oshlack et al. 2001; Zhou \etal\ 2003, 2005; see also
Maccarone \etal\ 2005).  As discussed by Zhou \etal\ (2003, 2005),
this is not entirely unexpected, since NLS1s are unlikely to be a
homogeneous class.  To the extent that the broad-line region, or at
least the portion emitting \hbeta, may have a disklike geometry, it
seems unavoidable that {\it some}\ NLS1s must be sources with
intrinsically large velocity dispersions viewed preferentially face-on
as opposed to having low \mbh\ (see also Bian \& Zhao 2004).
Perhaps more baffling is why such a small fraction of NLS1s are radio-loud
as compared to the general quasar population.

\begin{figure*}[t]
\begin{center}
\vskip +0.1truein
\psfig{file=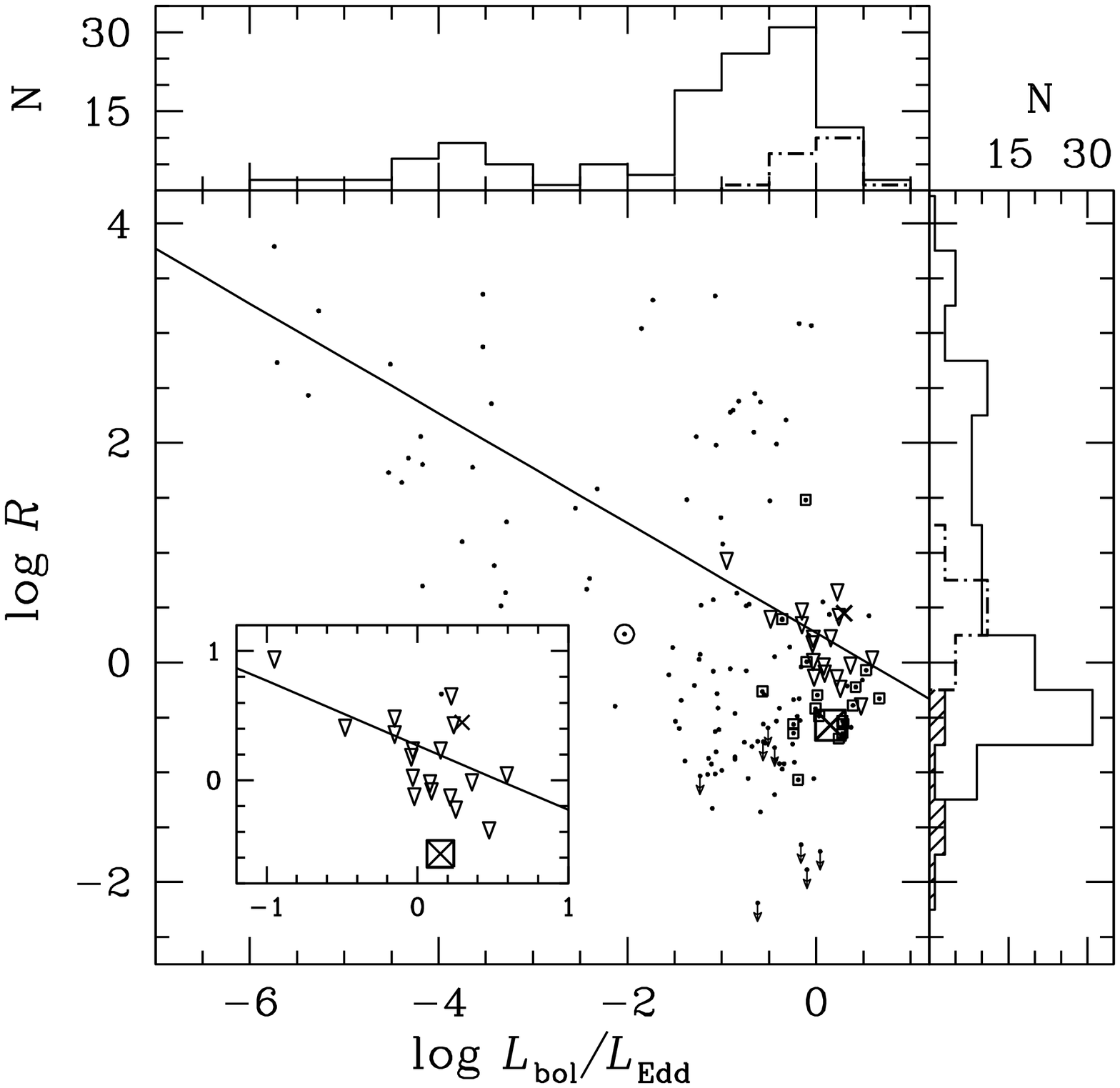,width=0.8\textwidth,keepaspectratio=true,angle=0}
\vskip -3mm 
\figcaption[]{ 
The trend of decreasing \rkel\ with
increasing \lledd.  The upper limits from this study are shown as
triangles, GH10 is shown with a cross, and our stacked upper limit is
plotted as the boxed cross.  NGC 4395 is highlighted with a semifilled
circle.  For comparison, we include the 87 PG quasars with $z < 0.5$,
using radio continuum data from Kellermann \etal\ (1989) and optical
continuum data from Neugebauer \etal\ (1987).  We obtained BH masses
with the method of Kaspi \etal\ (2000) using \fwhb\ from Boroson \&
Green (1992), and we estimated $L_{\mathrm{bol}}$ following Ho (2002).
PG objects with \fwhb\ $< 2000$~\kms\ (NLS1s) are boxed; they have a
mean \rkel\ of 0.54.  The data for the low-\lledd\ objects come from
Ho (2002), as does the overplotted best-fit line.  {\it Inset}: Our
current sample shown on an expanded scale for clarity.  {\it Top
histogram}: The solid-line histogram plots the distribution of \lledd\
for all the objects shown for comparison; the dash-dot histogram
highlights the 19 objects from this study.  {\it Right histogram}:
Same format as above, but for the distribution of \rkel.  Note that
all but one of the objects in the solid histogram are upper limits.
Upper limits in \rkel\ for the PG objects are shown with a shaded
histogram.  }
\vskip -5mm
\end{center}
\end{figure*}

\vspace{+1.2cm}
\section{Physical Interpretation}

This is the first high-resolution radio study of a uniformly selected
sample of AGNs with low mass and high Eddington ratio.  As a class,
these objects are radio-quiet down to very strong limits.  Our
observations thus provide new support for a model in which radio-loudness 
is anti-correlated with \lledd, as proposed, for example,~by Ho (2002).

AGNs at the lowest \lledd\ are observed to be generically radio-loud
(Ho 1999, 2002, 2005).  At low \lledd\ the conventional optically
thick, geometrically thin accretion disk (Shakura \& Sunyaev 1973) is
replaced by a radiatively inefficient accretion flow (RIAF; Quataert
2001; Narayan 2005 and references therein).  The RIAF itself
efficiently produces radio emission while simultaneously lacking the
optical-ultraviolet thermal emission from the standard accretion disk,
which acts to boost \rkel\ (Rees \etal\ 1982; Quataert \etal\ 1999; Ho
\etal\ 2000; Ho 2002).  At the same time, it has been shown in many
cases that a RIAF alone cannot account for all of the observed radio
emission, suggesting that an additional radio-emitting component, most
likely associated with a jet, is required (Yi \& Boughn 1999; Ulvestad
\& Ho 2001b; Yuan et al. 2002; Anderson et al. 2004; Falcke et
al. 2004).  On the opposite extreme, NLS1s, probably high-\lledd\ systems
as discussed above, are systematically radio-quiet (although
radio-loud quasars are exceptions; see below).  Broad absorption-line
(BAL) quasars are also predominantly radio-quiet (Stocke \etal\ 1992;
cf. Becker et al. 2000), and at least a subset of them (the
low-ionization subclass) are thought to be radiating at high \lledd\
(Meier 1996; Boroson 2002).  Principal component analysis of AGNs has
identified radio-quietness as a property that is statistically linked
with strong \feii\ emission and weak \oiii\ lines, a family of
properties, common to NLS1s and BAL quasars, that is thought to
correspond to high-\lledd\ systems (Boroson \& Green 1992;~Marziani
\etal\ 2001, 2003; Boroson 2002).

The trend of decreasing \rkel\ with increasing \lledd\ is shown
graphically in Figure 3 (cf. Fig. 5{\it b}\ in Ho 2002), using a
representative sample of AGNs with radio and optical luminosities and
BH mass estimates.  To extend the dynamic range in the plot, we have
augmented the sample from this study with (1) the $z<0.5$ PG quasars
from Boroson \& Green (1992) for which we can estimate virial BH
masses and (2) a sample of low-luminosity nuclei with BH masses
derived from direct dynamical modeling (Ho 2002).  Following standard
practice, we infer the bolometric luminosity directly from the optical
continuum luminosity by assuming a fixed bolometric correction (see,
e.g., discussion in Greene \& Ho 2004).  In low-luminosity sources
lacking reliable measurements of the optical AGN continuum, Ho (2002)
makes use of an additional empirical correlation between \hbeta\ and
continuum luminosity (Ho \& Peng 2001).  Our upper limits on \rkel,
and in particular the stronger limit derived from our stacking
analysis (large crossed box in Fig. 3), are consistent with
the observed trend, given the \lledd\ distribution of the sample.  To
properly sample the low-mass regime, we should include objects at low
\lledd\ as well.  While these sources are intrinsically faint even
when radiating at their Eddington luminosities, and thus difficult to
find, one example is known.  NGC 4395, with \mbh$\approx 10^4-10^5$
\msun, is believed to have a \lledd $\approx 0.01$ (Filippenko \&
Ho 2003).  It has \rkel\ $\approx 2.0$, which is significantly higher
than our stacked upper limit (\rkel\ = 0.27), and consistent with the
trend delineated in Figure 3.  Thus, the inverse correlation between
radio-loudness and accretion rate (\lledd) previously reported for
supermassive BHs ($10^6$ to few$\times 10^9$ \msun) appears to
continue to hold in the mass regime $10^4-10^6$ \msun.

To aid us in the physical interpretation of these results, we turn to
X-ray binaries (XRBs) that host BHs with \mbh \lax 10 \msun.  Because
the lower BH masses result in faster dynamical timescales, it is
possible to observe a single XRB in different accretion modes and link
them directly with changes in spectral properties. XRBs are in fact
found to occupy certain characteristic X-ray spectral states (see
McClintock \& Remillard 2005 for a review) corresponding to different
accretion regimes.  Typically, at low X-ray luminosity, during which
the accretion flow is thought to transform into a quasi-spherical RIAF
radiating at \lledd \lax 0.01 (Esin \etal\ 1997), the spectrum is
nonthermal and hard (the ``low/hard'' state).  This state is usually
accompanied by persistent, flat-spectrum radio emission, correlated
with the X-ray emission and likely including a jet component
(e.g.,~Dhawan et al. 2000; Stirling \etal\ 2001; Corbel \etal\ 2003;
Fender \& Belloni 2004).  At higher X-ray luminosities (0.01 \lax\
\lledd \lax 0.3), a conventional optically thick, geometrically thin
disk (Shakura \& Sunyaev 1973) produces a spectrum dominated by
thermal, soft X-ray emission (the ``high/soft'' state), during which
the radio emission is quenched (Tananbaum \etal\ 1972; Fender \etal\
1999, 2004; Gallo et al. 2003; Tigelaar \etal\ 2004).  Finally, for
systems with the highest luminosities (\lledd \gax 0.3), in which the
accretion flow might take the form of a radiation pressure-dominated
slim disk (Abramowicz et al. 1988), the X-ray spectrum can be
dominated by a very steep power-law component (the ``very-high'' or
``SPL'' state).  The very-high state itself is radio-quiet, but
transitions from the very-high to the high state may be accompanied by
optically thin ejection events in the radio (e.g.,~Corbel \etal\ 2001;
Hannikainen \etal\ 2001; Fender \etal\ 2004).

Since jet physics is basically scale-free (Heinz \& Sunyaev 2003), a number of 
authors have drawn a direct analogy between the spectral states of XRBs and 
AGNs (e.g.,~Meier 2001; Maccarone et al. 2003; Merloni et al. 2003; Falcke 
et al. 2004; Jester 2005).  The most thorough comparison has been made for 
low-luminosity AGNs, whose characteristic radio-loudness, hard X-ray spectra, 
and radiative inefficiency bear close resemblance to the properties of XRBs 
in their low/hard state (Ho 2005).  The RIAF at low accretion rates may be 
particularly conducive to jet production because its vertical thickness 
supports a strong poloidal magnetic field (Meier 2001), in addition to 
providing a population of weakly bound or unbound particles that are unstable 
toward outflow (Blandford \& Begelman 1999).  

Extending the comparison to the high and very-high states, however,
presents a greater challenge.  This is in part due to the somewhat
nebulous luminosity criterion various authors use to differentiate the
high state from the very-high state, and to the fact that the current
estimates of bolometric luminosities and BH masses in AGNs are still
rather uncertain.  NLS1s, with their characteristically soft X-ray
spectra, are often thought to be the direct analog of XRBs in the high
state (Pounds \etal\ 1995).  On the other hand, many NLS1s (e.g.,
Collin \& Kawaguchi 2004), though certainly not all (e.g., NGC 4395),
evidently have Eddington ratios---when taken at face value---that
exceed 0.3, the formal upper limit for the stability of a Shakura \&
Sunyaev disk.  If we take this theoretically motivated
criterion as a definition for the very-high state, then most
well-studied NLS1s, including the sample of Greene \& Ho (2004)
highlighted here, would technically qualify as being in the very-high
state, rather than in the high state.  Regardless of which choice one adopts 
for NLS1s, our study emphasizes a distinctive feature of NLS1s as a class:
they appear to be abnormally radio-quiet.  If the currently estimated
bolometric luminosities and BH masses are robust, the implication is
that jet production in AGNs is suppressed when their Eddington ratios
approach or exceed 1.  This empirical result for NLS1s is consistent
with the tendency for radio emission to be quenched in XRBs in their
high and very-high states.  The quenching of radio emission in the
high state may be related to the weakness of the poloidal component of
the magnetic field, central to driving a jet, in a geometrically thin
disk (Livio et al. 1999; Meier 2001).  As mentioned above, XRBs
undergoing very-high state transitions often exhibit rapid radio
flares, perhaps generated from the relativistic ejection of the corona
as the optically thick accretion disk moves inward (Fender et
al. 2004).  Since the general quasar population is thought to radiate
at high \lledd, it has been suggested that the radio-loud quasar
population consists of objects that have recently undergone a state
transition involving a relativistic ejection event (Marscher \etal\
2002; Gallo \etal\ 2003; Jester 2005).  In that case, it is puzzling
that NLS1s have such a low fraction (0\%--6\%) of
radio-loud objects as compared to quasars (10\%--20\%).  BAL quasars are
similar in this respect (e.g.,~Stocke \etal\ 1992).  Apparently at
the highest accretion rates, even relativistic ejection events are
uncommon.  This result warrants further theoretical exploration.

\section{Summary}

We present a high-resolution study of radio emission from a
well-defined sample of low-mass active galaxies radiating at a high
fraction of their Eddington luminosity (\lledd).  Only a single object
(GH10) in our sample of 19 is detected, with a radio power of 8.5
$\times 10^{21}$ \whz, while we place $3~\sigma$ upper limits of $3 \times
10^{20}$ to $8 \times 10^{21}$ \whz\ for the nondetections.  All the
objects have very low ratios of radio to optical flux (\rkel).  GH10
has an \rkel\ of 2.8, and a stacked image of the remaining observations
provides an upper limit of \rkel\ $\leq 0.27$.  While this study
represents one of the few deep, high-resolution radio surveys of a
uniform sample of narrow-line Seyfert 1 galaxies (NLS1s), a review of
the literature reveals that NLS1s are generically radio-quiet.
In contrast, low-\lledd\ sources are ubiquitously radio-loud.
Therefore, our observations provide renewed support for an inverse
relation between \lledd\ and radio-loudness.  Further support for this
picture is provided by black holes in X-ray binaries (XRBs), whose
radio emission is quenched when they radiate at high ($\geq 0.01$)
fractions of their Eddington luminosity.  This same analogy has been
used to suggest that radio-loud quasars may be undergoing a
relativistic ejection event similar to those observed when XRBs
transition from the very-high state.  In that case, it is quite
intriguing that objects at the highest accretion rates, namely 
NLS1s and BAL quasars, have abnormally low radio-loud fractions, a 
finding for which we currently have no clear explanation.

\acknowledgements 

We are grateful for useful discussions with J. McClintock, 
R. Narayan, and J.-M.~Wang.  
We thank the staff at the Very Large Array for hosting a very
productive visit for J.~E.~G. and L.~C.~H.  L.~C.~H. acknowledges support
by the Carnegie Institution of Washington and by NASA grants from the
Space Telescope Science Institute (operated by AURA, Inc., under NASA
contract NAS5-26555).  


\end{document}